\documentclass[pra,aps,onecolumn,showpacs,tightenlines]{revtex4}
\usepackage{graphics,bm}
\usepackage{graphicx}
\def\bmath#1{\mbox{\boldmath$#1$}}
\newcommand{\beq}{\begin{equation}}
\newcommand{\eeq}{\end{equation}}
\newcommand{\bqa}{\begin{eqnarray}}
\newcommand{\eqa}{\end{eqnarray}}
\protect

\begin{document}

\title{Vortex Stability Near the Surface of a Bose-Einstein Condensate}
\author{U. Al Khawaja}
\affiliation{ \it Physics Department, United Arab Emirates
University, P.O. Box 17551, Al-Ain, United Arab Emirates.}

\date{\today}

\begin{abstract}
We investigate energetic stability of vortices near the surface of
a Bose-Einstein condensate. From an energy functional of a
rotating Bose-Einstein condensate, written in terms of variables
local to the surface, and a suitable trial wavefunction we
calculate the energy of a moving vortex. The energetic stability
of the vortex is investigated in terms of the rotation frequency
of the confining potential. The critical frequency at which the
vortices enter the condensate is calculated and compared with the
experiment where a reasonable agreement is obtained.
\end{abstract}


 \maketitle

\section{Introduction}
\label{introduction} Since the experimental achievement of
Bose-Einstein condensation in confined dilute gases and the
observation of vortex lattices
\cite{mad,mad2,haj,ram,abo,abo2,hod}, much interest has been
devoted to the formation, stabilization, and dynamics of vortices
\cite{mad,mad2,butt,castin,dalfovo,feder,fetter,fetter2,matt,mad3,anatoly,james1,james2}.
Vortices have been created in Bose-Einstein condensates using two
different techniques. The first is a phase imprinting technique
\cite{matt} and the second is a rotating harmonic trap technique
\cite{mad,haj,abo,hod}. In the latter technique, a rotating laser
beam superposed on the magnetic trap creates an anisotropic
confining potential that stirs the condensate. Energetically
stable vortices are generated if the rotation frequency $\Omega$
of the harmonic trap is greater than a certain
critical frequency $\Omega_c$ \cite{baym,dalfovo2,anatoly,feder2}.

As a mechanism for vortex nucleation, it was suggested that
vortices are created from the excitation of low-energy surface
modes of the Bose-Einstein condensate
\cite{dalfovo3,isos,james1,mur,tsu}. For rotational frequencies
$\Omega<\Omega_c$, moving vortices are created in the ultra-dilute
region of the condensate. Their effect is observed at the surface
of the condensate only as a surface wave
\cite{gardiner,james1,james2,usama}. Upon increasing the rotation
frequency, the vortices approach the surface of the condensate. At
the critical frequency $\Omega_c$, the vortices {\it enter} the
condensate, i.e., nucleate. This was supported by Penckwitt {\it
et. al} \cite{gardiner} with a simulation to the growth of the
condensate from a rotating thermal cloud. It was also supported by
Anglin \cite{james1,james2} who showed, using a boundary layer
approach, that the energy barrier to vortex penetration disappears
at the Landau critical velocity for surface modes.

This picture for vortex nucleation will be the main focus of the
present paper. Using a (rather simple) variational approach to the
problem, we investigate the energetic stability of a vortex moving
near the surface of the condensate. We verify the above picture by
calculating the energy in terms of the distance of the vortex core
from the surface $x_0$, the rotation frequency $\Omega$, and the
center-of-mass speed of the core of the vortex $v_0$. We also
calculate the critical frequency $\Omega_c$ at which the vortex
enters the condensate.

In terms of variables local to the surface of large condensates,
the surface can be approximated by a plane and the harmonic
trapping potential can be approximated by a linear potential
\cite{emil,usama}. We start by writing the Gross-Pitaevskii energy
functional of a rotating condensate in this planar geometry. Using
this energy functional with an appropriate trial wavefunction we
calculate the energy of a vortex moving in a direction parallel to
the surface of the condensate. Then we minimize the energy with
respect to the position of the core of the vortex. The resulting
equilibrium position of the vortex core is a function of the
rotation frequency, the speed of the vortex, and the distance
between the vortices in the case when more than a single vortex
exists. This allows for a detailed investigation of the stability
of the vortex in terms of these parameters.

The rest of the paper is organized as follows. In
Sec.~\ref{functional_sec} we introduce the energy functional of a
rotating condensate in the planar geometry. In Sec.~\ref{sec2} we
use this energy functional to calculate the energy of a moving
vortex as described above. In Sec.~\ref{stability} we investigate
the stability of the vortex in terms of the parameters involved.
We end in Sec.~\ref{conclusion} by a summary and discussion of our
results.

\section{Energy Functional of the Surface of a rotating
condensate} \label{functional_sec} In this section we write the
Gross-Pitaevskii equation of a rotating condensate in terms of
parameters local to the surface of the condensate. Then we
construct an energy functional that corresponds to this equation.

The time-dependent Gross-Pitaevskii equation, that describes the
behavior of the order parameter $\psi({\bf r},t)$, is given by
\begin{equation}
\left[-{\hbar^2\over2m}\bmath{\nabla}^2+V({\bf r},t)+g|\psi({\bf
r},t)|^2-\mu\right]\psi({\bf r},t)=i\hbar{\partial\over\partial
t}\psi({\bf r},t)\;. \label{gp}
\end{equation}
Here $V({\bf r},t)$ is a time-dependent harmonic trapping
potential with time-dependence corresponding to rotating the axes
of the trapping potential. The effective two-particle interaction
$g$ is proportional to the $s$-wave scattering length $a$
according to $g=4\pi a\hbar^2/m$, where $m$ is the mass of an
atom, and $\mu$ is the equilibrium chemical potential.

We consider the trap axes in the $x$- and $y$-directions to be
rotated counter clockwise around the $z$-axis with angular
frequency $\Omega$. The time-dependence of the trapping potential
can be removed by writing the Gross-Pitaevskii equation in the
rotating frame
\begin{equation}
\left[-{\hbar^2\over2m}\bmath{\nabla}^2+V({\bf r})+g|\psi({\bf
r},t)|^2-{\bf\Omega\cdot}{\bf L}-\mu\right]\psi({\bf
r},t)=i\hbar{\partial\over\partial t}\psi({\bf r},t)\;,
\label{gp_rotating}
\end{equation}
where ${\bf L}$ is the angular momentum operator.
The trapping
potential, which for simplicity is taken to be isotropic, is given
by
\begin{equation}
V({\bf r})={1\over2}m\omega_0^2r^2\;, \label{v}
\end{equation}
where $\omega_0$ is the characteristic frequency of the trap.
Noting that ${\bf\Omega\cdot}{\bf L}=-i\hbar{\bf\Omega}\cdot({\bf
r}\times{\bf \bmath{\nabla}})=-i\hbar({\bf\Omega}\times{\bf
r})\cdot{\bf\bmath{\nabla}}$, the Gross-Pitaevskii equation takes
the form
\begin{equation}
\left[-{\hbar^2\over2m}\bmath{\nabla}^2+V({\bf r})+g|\psi({\bf
r},t)|^2+i\hbar({\bf\Omega}\times{\bf
r})\cdot{\bf\bmath{\nabla}}-\mu\right]\psi({\bf
r},t)=i\hbar{\partial\over\partial t}\psi({\bf r},t)\;.
\label{gp_rotating2}
\end{equation}
Next, we express the Gross-Pitaevskii equation in terms of the
surface local coordinate $\bf r^\prime$ and surface local velocity
$\bf v^\prime$ via a position transformation
\begin{equation}
{\bf r}=R{\bf \hat r}+{\bf r^\prime}\;, \label{ptrans}
\end{equation}
where $R=\sqrt{2\mu/m\omega_0^2}$ is the Thomas-Fermi radius of
the condensate. The Gross-Pitaevskii equation takes the form
\begin{equation}
\left[-{\hbar^2\over2m}\bmath{\nabla}^2_{\bf r^\prime}+V(R{\bf\hat
r}+{\bf r^\prime})+g|\psi({\bf r^\prime
},t)|^2+i\hbar({\bf\Omega}\times{\bf
r^\prime})\cdot{\bf\bmath{\nabla}_{\bf r^\prime }}
+i\hbar({\bf\Omega}\times R{\bf\hat
r})\cdot{\bf\bmath{\nabla}_{\bf r^\prime }}-\mu\right]\psi({\bf
r^\prime },t) =i\hbar{\partial\over\partial t}\psi({\bf
r^\prime},t)\;. \label{gp_rotating2.5}
\end{equation}
The term $i\hbar({\bf\Omega}\times R{\bf\hat
r})\cdot{\bf\bmath{\nabla}_{\bf r^\prime }}$ can be eliminated by
a Galilean transformation to a frame of reference that is moving
with a speed ${\bf\Omega}\times R{\bf\hat R}$ \cite{fetter5}.
Thus, the Gross-Pitaevskii equation becomes
\begin{equation}
\left[-{\hbar^2\over2m}\bmath{\nabla}^2_{\bf r^\prime}+V(R{\bf\hat
r}+{\bf r^\prime})+g|\psi({\bf r^\prime
},t)|^2+i\hbar({\bf\Omega}\times{\bf
r^\prime})\cdot{\bf\bmath{\nabla}_{\bf r^\prime }}
-\mu\right]\psi({\bf r^\prime },t) =i\hbar{\partial\over\partial
t}\psi({\bf r^\prime},t)\;. \label{gp_rotating3}
\end{equation}
For distances close to the surface of the condensate, such that
$r^\prime\ll R$, the trapping potential can be approximated by a
linear function as \cite{emil}
\begin{eqnarray}
V(R{\bf\hat r}+{\bf r^\prime}) &\simeq&
V(R)+{\bf\bmath{\nabla}_{\bf r^\prime}}V(R{\bf\hat r}+{\bf
r^\prime})
|_{\bf r^\prime=0}\cdot{\bf r^\prime} \nonumber\\
&=& V(R)+m\omega_0^2R\,{{\bf \hat r}\cdot}{\bf
r^\prime}\nonumber\\
&=& V(R)+Fx \;,\label{linear}
\end{eqnarray}
where we have defined the {\it force constant} $F=m\omega_0^2R$
and the surface local coordinate $x=\hat{\bf r}\cdot\bf r^\prime
$, which is normal to the surface of the condensate. The
coordinates of the $y$ and $z$-components of $\bf r^\prime$ are
therefore parallel to the surface. In this plane geometry the
Gross-Pitaevskii equation takes the form
\begin{eqnarray}
\left[-{\hbar^2\over2m}\bmath{\nabla}^2_{\bf
r^\prime}+F\,x+g|\psi({\bf r^\prime
},t)|^2+i\hbar({\bf\Omega}\times{\bf
r^\prime})\cdot{\bf\bmath{\nabla}_{\bf r^\prime }}\right]\psi({\bf
r^\prime },t)=i\hbar{\partial\over\partial t}\psi({\bf
r^\prime},t)\;. \label{gp_rotating4}
\end{eqnarray}
The first and the last terms on the left hand side of this
equation can be combined as a complete square, which results in
the following form for the Gross-Pitaevskii equation
\begin{eqnarray}
\left[-{\hbar^2\over2m}\left(\bmath{\nabla}_{\bf
r^\prime}-{im\over\hbar}{\bf\Omega}\times{\bf r^\prime
}\right)^2+F\,x+g|\psi({\bf r^\prime
},t)|^2-{1\over2}m({\bf\Omega}\times{\bf r^\prime
})^2\right]\psi({\bf r^\prime },t)=i\hbar{\partial\over\partial
t}\psi({\bf r^\prime},t)\;. \label{gp_rotating5}
\end{eqnarray}
This equation can be derived by differentiating the following
lagrangian with respect to $\psi^*({\bf r^\prime},t)$
\begin{equation}
L[\psi,\psi^*]= \int{d{\bf r^\prime}{i\hbar\over2}\left(
\psi^*{\partial\psi\over\partial t}
-\psi{\partial\psi^*\over\partial t} \right)-E[\psi,\psi^*]}\,,
\label{lagrangian}
\end{equation}
where $E[\psi,\psi^*]$ is given by
\begin{eqnarray}
E[\psi,\psi^*]=\int{d{\bf
r^\prime}\left[{\hbar^2\over2m}\left|\left(\bmath{\nabla}_{\bf
r^\prime}-{im\over\hbar}{\bf\Omega}\times{\bf r^\prime
}\right)\psi({\bf r^\prime },t)\right|^2+F\,x|\psi({\bf r^\prime
},t)|^2+{1\over2}g|\psi({\bf r^\prime
},t)|^4-{1\over2}m({\bf\Omega}\times{\bf r^\prime })^2|\psi({\bf
r^\prime },t)|^2\right]}\;. \label{energy}
\end{eqnarray}

\section{energy of a moving vortex near the surface of the condensate}
\label{sec2}

\begin{figure}[htb]
\begin{center}
\includegraphics[width=10.cm]{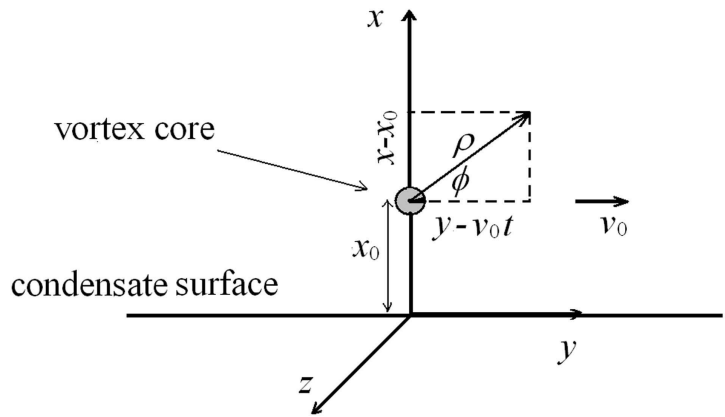}
\end{center}
\caption[a]{
Figure showing the vortex core located at a distance $x_0$ from
the Thomas-Fermi surface of the condensate and is moving with a
speed $v_0$.} \label{fig1}
\end{figure}

We consider a vortex with axis parallel to the surface of the
condensate, say parallel to the $z$-direction. The core of the
vortex is located at a distance $x=x_0$ from the Thomas-Fermi
surface of the condensate and is moving in the positive
$y$-direction with a speed $v_0$. This is depicted in the
schematic figure Fig.~\ref{fig1}. For simplicity, we consider a
vortex with a small core size such that the equilibrium condensate
density can be considered constant over distances of the order of
the core size.

The wavefunction of the condensate that contains such a vortex can
thus be written as the product of the wave functions of the
condensate that contains no vortex and the wave function of the
vortex with an appropriate phase factor. The phase factor should
take into account both the superfluid and the center of mass
velocities of the vortex. Specifically, we write the wave function
of the vortex as
\begin{equation}
\psi({\bf
r^\prime},t)=\sqrt{n_0(x)}\chi(x-x_0,y-v_0t)\exp{[i\Phi(x-x_0,y-v_0t)]}\;.
\label{psitrial}
\end{equation}
Here, $n_0(x)=|\psi_0(x)|^2$ is the density profile of the {\it
vortex-free} condensate at equilibrium. The equilibrium
wavefunction $\psi_0(x)$ obeys Eq.~(\ref{gp_rotating5}) after
setting $\bf\Omega$ and $\partial\psi/\partial t$ to zero, namely
\begin{equation}
\left[-{\hbar^2\over2m}{d^2\over
dx^2}+Fx+g|\psi_0(x)|^2\right]\psi_0(x)=0\;. \label{gpx}
\end{equation}

The normalized density profile of the vortex is given by
$\chi^2(x-x_0,y-v_0t)$, where we assume for  $\chi(x-x_0,y-v_0t)$
the trial function \cite{fetter_old}
\begin{equation}
\chi(x-x_0,y-v_0t)={\rho\over\sqrt{2\xi^2+\rho^2}}\;, \label{chi2}
\end{equation}
whith $\rho=\sqrt{(x-x_0)^2+(y-v_0t)^2}$, and $\xi=1/\sqrt{8\pi a
n_0(x_0) }$ is the coherence length of the condensate calculated
at $x_0$.

The phase of the vortex $\Phi(x-x_0,y-v_0t)$ must satisfy
\begin{eqnarray}
 {\hbar\over
m}{\bf\bmath{\nabla}_{r^\prime}}\Phi&=&{\hbar\over m\rho}
\bmath{\hat\phi}+v_0{\bf \hat y}\nonumber\\
&=&{\hbar\over m\rho}\cos\phi\;{\bf\hat x}+(v_0-{\hbar\over
m\rho}\sin\phi){\bf\hat y}\;, \label{phase}
\end{eqnarray}
where $\rho$ and $\phi$ are the polar coordinates in the $x$-$y$
plane with the origin being at the core of the vortex as shown in
Fig.~\ref{fig1}. This equation expresses the fact that the
velocity has two components: the superfluid velocity
$\hbar{\bmath{ \hat\phi}}/m\rho$ associated with the vortex, and
the center-of-mass velocity $v_0{\bf\hat y}$.

In Appendix A, we show that using the trial function
Eq.~(\ref{psitrial}) in the energy functional Eq.~(\ref{energy}),
the energy of the vortex per unit length along the axis of the
vortex takes the form
\begin{eqnarray}
E&=&n_0(x_0)\int{d{\bmath\rho}\left[{\hbar^2\over2m}
\left(({\bmath\nabla}\chi)^2+{\chi^2\over\rho^2}\right)
+{1\over2}g\,n_0(x_0)\chi^4\right]}
\nonumber\\
&+& n_0(x_0) \left[
{1\over2}mv_0^2+\hbar\Omega-m\Omega\,x_0v_0+F\,x_0
\right]\int{d{\bmath\rho}\chi^2}\;. \label{energy2}
\end{eqnarray}

The quantity of interest is the extra energy associated with the
presence of the vortex \cite{book}. Therefore, we have to subtract
the vortex energy given by the last equation from the energy of a
vortex-free system that contains the same number of atoms as the
system with the vortex. The energy density of the vortex-free
system is only that of the mean-field interaction, the trapping
potential, and the energy associated with the rotation, namely
\begin{equation}
E_{\rm
vf}={1\over2}gn^2_0(x_0)+F\,x_0n_0(x_0)+\hbar\Omega\,n_0(x_0)\,,
\label{evf}
\end{equation}
where the subscript ``vf'' in $E_{\rm vf}$ denotes vortex free. To
fulfil the requirement that the number of atoms $\nu$ in both
systems must be the same, we calculate the energy of both systems
within a cylindrical region of axis normal to the $x$-$y$ plane
and circular cross-section of radius $b$. The radius $b$ must be,
on the one hand, larger than several coherence lengths $\xi$ in
order to incorporate all density variations associated with the
presence of the vortex. On the other hand, $b$ must be much
smaller than the length over which the density changes
significantly in order not to violate the local-density
approximation used to derive calculate the energies
Eqs.~(\ref{energy2}) and (\ref{evf}). The number of atoms per unit
length of the axis of the cylinder in the system that contains the
vortex can be expressed as
\begin{eqnarray}
\nu&=n_0(x_0)&\int_0^b{2\pi\rho\,d\rho\,\chi^2}\nonumber\\
&=&\pi b^2n_0(x_0)-n_0(x_0)\int_0^b{2\pi}\rho\,d\rho
\left(1-\chi^2\right)\;. \label{nu}
\end{eqnarray}
Using this number to calculate the average density as
$n_0(x_0)=\nu^2/\pi\,b^2$, the energy per unit length of the
vortex-free system is given by
\begin{equation}
E_{\rm vf}\simeq{1\over2}\pi\,b^2gn_0^2(x_0)-gn_0^2(x_0)
\int_0^b{2\pi\rho\,d\rho(1-\chi^2)}+\left(Fx_0n_0(x_0)+\hbar\Omega
n_0(x_0)\right)\int{2\pi\rho\,d\rho\chi^2}. \label{evf2}
\end{equation}
In this equation we ignore terms that are second order in
$1-\chi^2$ in the mean-field energy \cite{book}.

Subtracting Eq.~(\ref{evf2}) from Eq.~(\ref{energy2}), we obtain
the energy associated with the presence of the vortex per unit
length of its axis:
\begin{eqnarray}
E_{\rm v} &=&n_0(x_0)\int{d{\bmath\rho}\left[{\hbar^2\over2m}
\left(({\bmath\nabla}\chi)^2+{\chi^2\over\rho^2}\right)
+{1\over2}g\,n_0(x_0)(\chi^2-1)^2\right]}
\nonumber\\
&+& n_0(x_0) \left[
{1\over2}mv_0^2-m\Omega\,x_0v_0\right]\int{d{\bmath\rho}\chi^2}
+\hbar\Omega\int{d\,{\bmath\rho}(\chi^2-1)}+Fx_0n_0(x_0)\int{d\,{\bmath\rho}(\chi^2-1)}\;.
\label{energy_last}
\end{eqnarray}
We recognize the first line of this equation as the {\it internal}
energy of the vortex \cite{book}.

\section{Stability analysis}
\label{stability} In this section we consider an infinite string
of equidistant moving vortices. The string is parallel to the
surface of the condensate, say along the $y$-axis, and the
vortices move in that direction with a constant speed $v_0$. The
distance between two adjacent vortices $\lambda$ is large enough
such that the cores do not overlap. In this case, the energy of
the string of vortices per unit length of the string is just the
energy of one vortex given by Eq.~(\ref{energy_last}) times the
number of vortices per unit length. Therefore, it is sufficient to
perform the analysis by calculating the energy of a single vortex.
This energy will be a function of $\lambda$ that corresponds to a
certain number of vortices per unit length.

In the following, we investigate the dependence of the equilibrium
position of the vortex on the rotation frequency $\Omega$, the
distance between two adjacent vortices $\lambda$, and the vortex
speed $v_0$.

For convenience, we rewrite Eq.~(\ref{energy_last}) in terms of
scaled quantities as follows:
\begin{equation}
\tilde{E}_{\rm v}=\tilde{\epsilon}_{\rm
v}+\nu_1\tilde{n}_0(\tilde{x}_0)(
\tilde{v}_0^2-2\tilde{\Omega}\,\tilde{x}_0\tilde{v}_0)
+\nu_2\tilde{n}_0(\tilde{x}_0)(2\tilde{\Omega}+\tilde{x}_0)\;,
\label{energy_last2}
\end{equation}
where
$\nu_1=2\pi\int_0^{\tilde{b}}{\tilde{\rho}d{\tilde{\rho}}\chi^2}$,
$\nu_2=2\pi\int_0^{\tilde{b}}{\tilde{\rho}d{\tilde{\rho}}(\chi^2-1)}$,
and
\begin{equation}
\tilde{\epsilon}_{\rm v}
=2\pi\tilde{n}_0(\tilde{x}_0)\int_0^{\tilde{b}}{\tilde{\rho}d\tilde{\rho}}\left[
\left(({\bmath\nabla}\chi)^2+{\chi^2\over\tilde{\rho}^2}\right)
+{1\over2}\tilde{n}_0(\tilde{x}_0)(\chi^2-1)^2\right]\;,
\label{ev_internal}
\end{equation}
is the internal energy of the vortex. Here, the gradient operator
is with respect to $\tilde{\bmath\rho}$. Length is scaled to the
characteristic length at the surface of the condensate $\delta$
defined by $\hbar^2/2m\delta^2=F\delta$. The vortex velocity $v_0$
is scaled to $\hbar/m\delta$, the angular frequency $\Omega$ is
scaled to $\hbar/m\delta^2$, the density $n_0$ is scaled to $8\pi
a\delta^2$, and the energies $E_{\rm v}$ and $\epsilon_{\rm v}$
are scaled to $(\hbar^2/2m\delta^2)/(8\pi
a/\delta)=((2R/a_0)^{1/3}/16\pi)\hbar\omega_0$. In these
dimensionless variables, Eq.~(\ref{gpx}), takes the form
\begin{equation}
\left[-{d^2\over
d\tilde{x}^2}+\tilde{x}+\tilde{n}_0(\tilde{x}_0)\right]\sqrt{\tilde{n}_0(\tilde{x}_0)}=0\;.
\label{gpx_scaled}
\end{equation}
Furthermore, the vortex profile $\chi$, defined in
Eq.~(\ref{chi2}), takes the form
\begin{equation}
\chi={\tilde{\rho}\over\sqrt{2/\tilde{n}_0(\tilde{x}_0)+\tilde{\rho}^2}}\;.
\label{chi2_scaled}
\end{equation}
Since the distance between two adjacent vortices is
$\tilde{\lambda}$, the upper limit of the integrations in the
expressions for $\nu_1$, $\nu_2$, and Eq.~(\ref{ev_internal}) is
taken as $\tilde{b}=\tilde{\lambda}/2$.

We start by minimizing the vortex energy $\tilde{E}_{\rm v}$ with
respect to the vortex linear speed $\tilde{v}_0$. This gives
simply $\tilde{v}_0=\tilde{x}_0\tilde{\Omega}$, which is similar
to the relation between the center of mass speed and the angular
frequency of a rolling rigid wheel of radius $\tilde{x}_0$.
Substituting this value for $\tilde{v}_0$ in
Eq.~(\ref{energy_last2}), the energy of the vortex becomes
\begin{equation}
\tilde{E}_{\rm v}=\tilde{\epsilon}_{\rm
v}-\nu_1\tilde{n}_0(\tilde{x}_0)\tilde{x}_0^2\,\tilde{\Omega}^2
+\nu_2\tilde{n}_0(\tilde{x}_0)(2\tilde{\Omega}+\tilde{x}_0)\;.
\label{energy_last3}
\end{equation}
The equilibrium position of the vortex $\tilde{x}_{\rm eq}$ is
obtained by minimizing the energy $\tilde{E}_{\rm v}$ with respect
to the vortex distance from the surface $\tilde{x}_0$, for a given
set of the parameters $\tilde{\Omega}$ and $\tilde{\lambda}$. The
main focus will be on the behavior of the equilibrium position
$\tilde{x}_{\rm eq}$ as a function of the parameters
$\tilde{\Omega}$ and $\tilde{\lambda}$. In particular, we are
interested in situations when the vortex {\it enters} the
condensate. The Thomas-Fermi surface, defined by $\tilde{x}_0=0$,
is to be taken as the boundary for considering the vortices inside
or outside the condensate. Vortices located in the region
$\tilde{x}_0>0$ are to be considered out of the condensate, while
those in the region $\tilde{x}_0<0$ are considered inside the
condensate.

To calculate the vortex energy $\tilde{E}_{\rm v}$, we first need
to solve Eq.~(\ref{gpx_scaled}) for $\tilde{n}_0(\tilde{x}_0)$. As
a first approximation we use the Thomas-Fermi approximation for
the condensate density $\tilde{n}_0(\tilde{x}_0)$. This is
obtained by neglecting the first term in Eq.~(\ref{gpx_scaled}),
namely
\begin{equation}
\tilde{n}_0(\tilde{x}_0)=\left\{
\begin{array}{cc} -\tilde{x}_0,&\tilde{x}_0<0\\
0,&\tilde{x}_0>0
\end{array}\right.
\;. \label{tf_density}
\end{equation}
The Thomas-Fermi approximation is accurate only in the deep region
of the condensate $\tilde{x}_0\ll0$. Thus, the results we obtain
here using this approximation will be accurate only in that
region. From Eq.~(\ref{chi2_scaled}) we see that the core size of
the vortex depends on the condensate density. This will complicate
the dependence of the energy on the vortex location. For
simplicity, we take the size of the core to be constant and is
equal to its value at the Thomas-Fermi surface. Substituting for
$\tilde{n}_0(\tilde{x}_0)$ from Eq.~(\ref{tf_density}) in
Eq.~(\ref{energy_last3}), the vortex energy $\tilde{E}_{\rm v}$
becomes a cubic function in $\tilde{x}_0$, which can be written as
\begin{equation}
\tilde{E}_{\rm v}=-(\alpha_1+2{\tilde\Omega}\nu_2){\tilde
x}_0+(\alpha_2-\nu_2){\tilde x}_0^2+\nu_1{\tilde\Omega}^2{\tilde
x}_0^3 \label{ev_tf}
\end{equation}
where
\begin{equation}
\alpha_1=2\pi\int_0^{\tilde{b}}{\tilde{\rho}d\tilde{\rho}}
\left(({\bmath\nabla}\chi)^2
+{\chi^2\over\tilde{\rho}^2}\right)\;, \label{alpha1}
\end{equation}
and
\begin{equation}
\alpha_2=2\pi\int_0^{\tilde{b}}{\tilde{\rho}d\tilde{\rho}}
{1\over2}(\chi^2-1)^2\;. \label{alpha2}
\end{equation}

To be able to extract the main features of this energy expression,
we point out some properties of the numbers $\nu_1$ and $\nu_2$
and the parameters $\alpha_1$ and $\alpha_2$. The number $\nu_1$
is always positive since it corresponds to the number of atoms
within a circle of radius $\tilde b$ centered at the axis of a
vortex that is embedded in a uniform background of density
$\tilde{n}_0=1$. The number $\nu_2$ is always negative since it is
the difference between the number of atoms in the vortex $\nu_1$
and the number of atoms in a vortex-free background.  The
parameter $\alpha_1$ corresponds to the kinetic energy due to the
gradient of the wavefunction of the vortex. The parameter
$\alpha_2$ corresponds to the mean-field interaction energy. It
should be also noted that Eq.~(\ref{ev_tf}) is defined only for
negative values of $\tilde{x}_0$.

With this in mind, we conclude from Eq.~(\ref{ev_tf}) that
$\tilde{E}_{\rm v}$ starts linearly for small $\tilde{x}_0$ with a
slope $-(\alpha_1+2\tilde{\Omega}\nu_2)$ that, depending on the
value of $\tilde{\Omega}$, can be either positive or negative. For
large $\tilde{x}_0$ the energy $\tilde{E}_{\rm v}$ behaves as
$\nu_1\tilde{\Omega}^2\tilde{x}_0^3$ which is large and negative.
If the initial slope $-(\alpha_1+2\tilde{\Omega}\nu_2)$ is
positive, $\tilde{E}_{\rm v}$ will have only a local maximum at
some value of $\tilde{x}_0$. If, on the other hand, the slope is
negative, $\tilde{E}_{\rm v}$ may have a local minimum in addition
to a local maximum. If the negative value of the slope is large
enough, the value of $\tilde{E}_{\rm v}$ at the local maximum can
vanish or even disappear. To be more explicit, we rewrite
Eq.~(\ref{ev_tf}) as
\begin{equation}
\tilde{E}_{\rm v}=\tilde{x}_0\left[
(\tilde{x}_0-x_1)(\tilde{x}_0-x_2) \right]\,,
\label{ev_tf2}
\end{equation}
where
\begin{equation}
x_1={-\alpha_2-\nu_2\over2\nu_1\tilde{\Omega}^2}
+{1\over2\nu_1\tilde{\Omega}^2}
\sqrt{(\alpha_2-\nu_2)^2+4\nu_1\tilde{\Omega}^2(\alpha_1+2\nu_2\tilde{\Omega})}
\label{x1}\,,
\end{equation}
and
\begin{equation}
x_2={-\alpha_2-\nu_2\over2\nu_1\tilde{\Omega}^2}
-{1\over2\nu_1\tilde{\Omega}^2}
\sqrt{(\alpha_2-\nu_2)^2+4\nu_1\tilde{\Omega}^2(\alpha_1+2\nu_2\tilde{\Omega})}
\label{x2}\,,
\end{equation}
are two of the three roots of Eq.~(\ref{ev_tf2}). The value of
$\tilde{E}_{\rm v}$ at the local maximum vanishes when $x_1=x_2$.
This can be satisfied if
\begin{equation}
(\alpha_2-\nu_2)^2+4\nu_1\tilde{\Omega}^2(\alpha_1+2\nu_2\tilde{\Omega})=0
\label{critomega}\,,
\end{equation}
which is the condition that gives the critical frequency
${\tilde\Omega}_c$ at which the height of the energy barrier for
vortex nucleation vanishes.

\begin{figure}[htb]
\begin{center}
\includegraphics[width=7.cm]{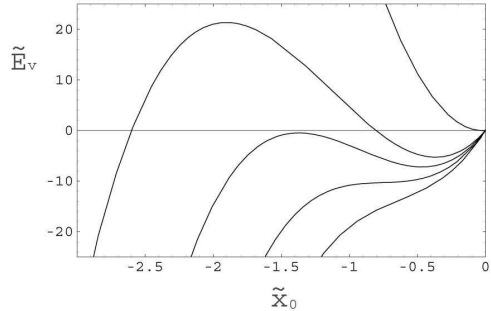}
\end{center}
\caption[a]{
Vortex energy $\tilde{E}_{\rm v}$ as a function of the distance
between the core of the vortex and the Thomas-Fermi surface
$\tilde{x}_0$. The energy is calculated using the Thomas-Fermi
approximation. Five curves are plotted for values of
$\tilde{\Omega}$ that equal, starting from the uppermost curve,
0.1, 0.45, 0.5, 0.55, 0.6. The value of $\tilde\lambda$ used is
21. The unit of energy is $((2R/a_0)^{1/3}/16\pi)\hbar\omega_0$
and the unit of distance is $\delta$. For a typical experiment
such as the one discussed at the end of Sec. IV, the unit of
energy equals $0.033\hbar\omega_0$, and the unit of distance
equals $0.6a_0$, where $a_0=\sqrt{\hbar/m\omega_0}$.}
\label{fig2}\end{figure}

The expression for ${\tilde E}_{\rm v}$ is plotted in
Fig.~\ref{fig2} as a function of ${\tilde x}_0$ for different
values of ${\tilde \Omega}$ and one value of $\tilde\lambda$. In
addition to the above-mentioned properties of the energy curve,
the figure shows that the height of the barrier decreases with
increasing $\tilde\Omega$. At a certain value of $\tilde\Omega$
the height of the barrier vanishes and the vortex can enter the
condensate. For the value of $\tilde \lambda$ used in this figure,
we find that, at $\tilde{\Omega}=0.5$, the height of the barrier
vanishes. This critical value of $\tilde{\Omega}$ is indeed equal
to the real root of Eq.~(\ref{critomega}).

Since the Thomas-Fermi approximation, used here, is not accurate
at the surface, we will not attempt to calculate the critical
frequency at which the vortex enters the condensate. This is to be
investigated next where we use the exact density profile of the
condensate.

For a more accurate treatment of the problem, we repeat the above
calculation using the numerical solution of Eq.~(\ref{gpx_scaled})
instead of the Thomas-Fermi one. We start by solving numerically
Eq.~(\ref{gpx_scaled}) for $\tilde{n}_0(\tilde{x}_0)$, and then
using this result in Eq.~(\ref{chi2_scaled}) to calculate $\chi$,
where the coherence length $\tilde\xi$ depends on
$\tilde{n}_0(\tilde{x}_0)$ and is not taken to be constant as in
the above calculation. These values of $\tilde{n}_0(\tilde{x}_0)$
and $\chi$ are then used to calculate $\tilde{E}_{\rm v}$ from
Eq.~(\ref{energy_last3}). Similar to the above result obtained
using the Thomas-Fermi approximation, we find that $\tilde{E}_{\rm
v }$ has a local minimum and a local maximum, and tends to
$-\infty$ as $\tilde{x}_0\rightarrow-\infty$.
\begin{figure}[htb]
\begin{center}
\includegraphics[width=7.cm]{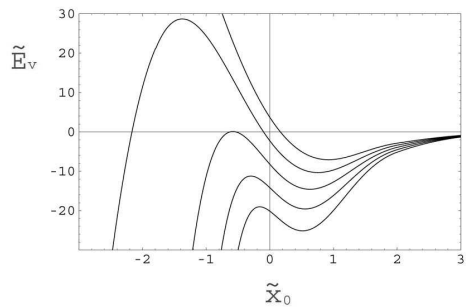}
\end{center}
\caption[a]{
Vortex energy $\tilde{E}_{\rm v}$ as a function of the distance
between the core of the vortex and the Thomas-Fermi surface
$\tilde{x}_0$. The energy is calculated numerically. Five curves
are plotted for values of $\tilde{\Omega}$ that equal, starting
from the uppermost curve, 0.0, 0.17, 0.34, 0.51, 0.68. The value
of $\tilde\lambda$ used is 21. Note that the height of the energy
barrier vanishes for ${\tilde\Omega}=0.34$ } \label{fig3}
\end{figure}
In Fig.~\ref{fig3} we plot the energy $\tilde{E}_{\rm v }$ as a
function of $\tilde{x}_0$ for different values of
$\tilde{\Omega}$. This figure is to be compared with Fig.~4 of
Ref.~\cite{james2} and Fig.~5 of Ref.~\cite{fetter2}. The main
feature of Figs.~\ref{fig2} and \ref{fig3}, which is that an
energy barrier for vortex nucleation exists and that the height of
the barrier decreases with increasing $\tilde\Omega$, agrees with
the above-mentioned two figures. However, a detailed comparison is
not possible for the following reasons. In Fig.~4 of
Ref.~\cite{james2} the energy is calculated in the {\it outer}
region of the condensate, which means that it is accurate only for
large distances away from the Thomas-Fermi surface. In Fig.~5 of
Ref.~\cite{fetter2} the energy is calculated using the
Thomas-Fermi approximation and therefore is accurate only inside
the condensate. The present calculation is valid in the region
near the surface of the condensate from both sides. Thus,
Fig.~\ref{fig3} presents a calculation of the energy in a region
where both calculations of Refs.~\cite{james2} and \cite{fetter2}
are not applicable.

\begin{figure}[htb]
\begin{center}
\begin{math}
\begin{array}{cc}
\includegraphics[width=7.cm]{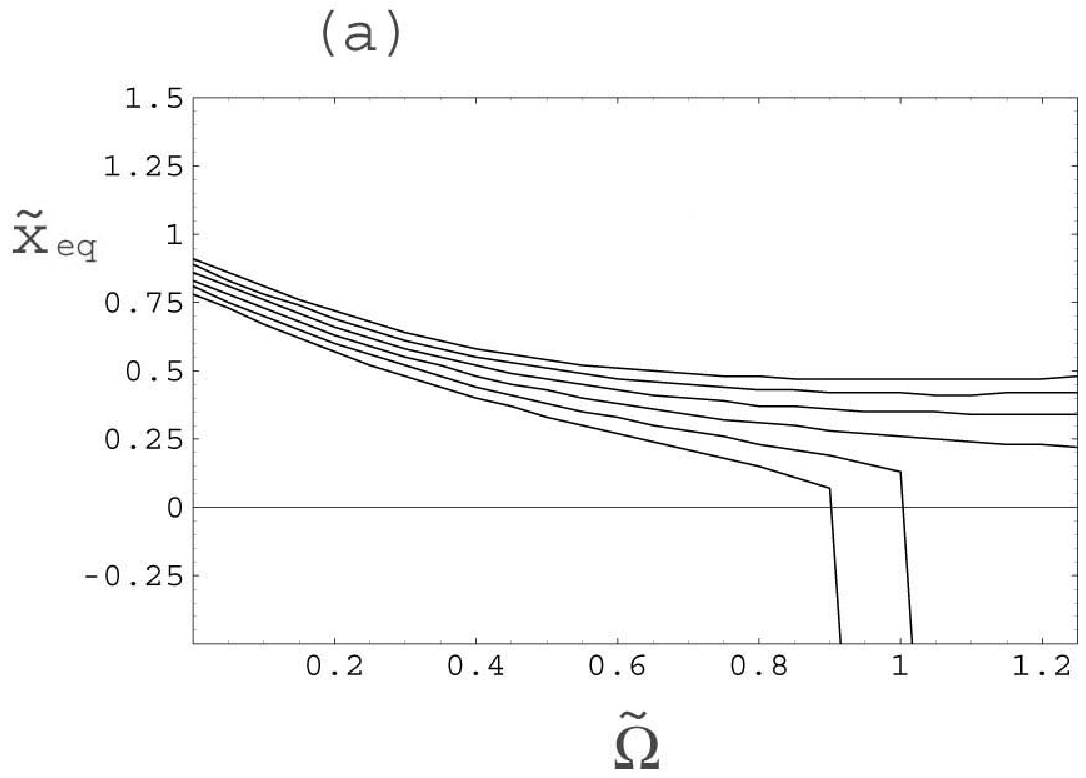}
&\\
\includegraphics[width=7.cm]{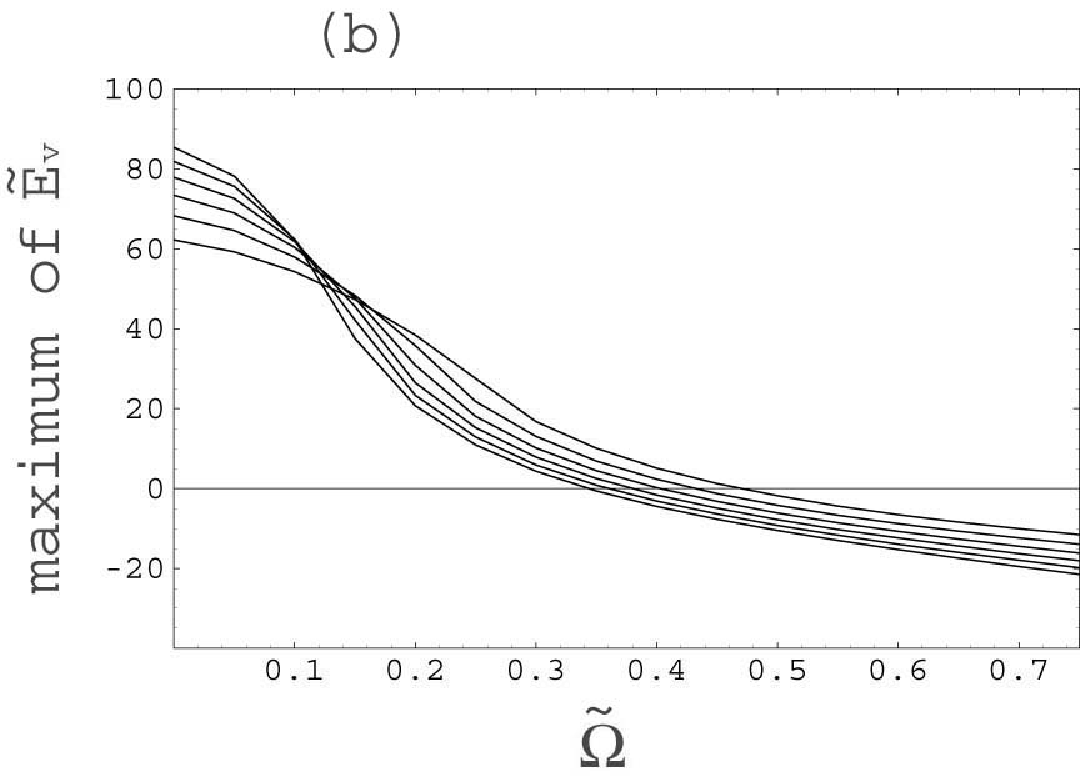}
&
\end{array}
\end{math}
\end{center}
\caption[a]{
a) The equilibrium position of the vortex core versus the rotation
frequency of the magnetic trap. These curves are calculated for
values of $\tilde\lambda$ that equal, starting from the lowest
curve, 10, 12, 14, 16, 18, 20. The sudden drop in two of the
curves appear because the local minimum of $\tilde{E}_{\rm v}$ is
no longer present. The local minimum and local maximum merge to
form a stationary point. b) The height of the energy barrier
calculated for values of $\tilde\lambda$ that equal, starting from
the rightmost curve, 10, 12, 14, 16, 18, 20. The units of energy
and distance are as those mentioned in Fig.~2, and the unit of
frequency is $(2R/a_0)^{2/3}\omega_0$, which for the experiment
discussed at the end of Sec.~IV equals $2.6\omega_0$.}
\label{fig4}
\end{figure}

In Fig.~\ref{fig4} we plot the position of the local minimum of
the energy and the height of the energy barrier as a function of
$\tilde{\Omega}$. Several curves are plotted corresponding to
different values of $\tilde{\lambda}$. Fig.~\ref{fig4}(a) shows
that the vortex approaches the surface when $\tilde{\Omega}$ is
increased. To enter the condensate, the energy barrier has to
vanish. Fig.~\ref{fig4}(b) shows that the energy barrier does
indeed vanish for a certain value of $\tilde{\Omega}$. This value
is to be considered as the critical value $\tilde{\Omega}_c$ at
which the vortex nucleates.

Now, we can compare our predictions for the critical frequency at
which the vortex nucleates with those observed experimentally. To
that end,  we need to express $\tilde{\Omega}_c$ in terms of the
harmonic trap frequency $\omega_0$ for a spherical condensate.
This is readily obtained by noticing that the characteristic
length scale at the surface of the condensate $\delta$ is related
to the Thomas-Fermi radius of the condensate through
$\delta=(\hbar^2/2mF)^{1/3}=(a_0/2R)^{1/3}a_0$, where
$a_0=\sqrt{\hbar/m\omega_0}$. Using this relation, $\tilde\Omega$
can be expressed in terms of $\omega_0$ as
${\tilde\Omega}=(\Omega/\omega_0)(a_0/2R)^{2/3}$, which simply
leads to $\Omega=(2R/a_0)^{2/3}\,{\tilde\Omega}\,\omega_0$.

In the experiment of the ENS group \cite{mad}, a single vortex was
created in about $1\times10^{5}$ atoms of a $^{87}$Rb condensate
at a rotating frequency $\Omega=2\pi\times147$ Hz. The magnetic
trap used was axially symmetric with axial trap frequency
$\omega_z=2\pi\times11.7$ Hz and radial trap frequency
$\omega_{\perp}=2\pi\times219$ Hz. The scattering length is taken
as $a=5.8$nm. For these parameters $R\approx2.1a_0$ and
$(2R/a_0)^{2/3}\approx2.6$, where we have used
$a_0=(\hbar/m(\omega_{\perp}^2\omega_z)^{1/3})^{1/2}$. For a
single vortex the value of $b$ is half the circumference of the
condensate, namely $b\approx2.1\pi\,a_0\approx10.5\delta$. Since
$\tilde{\lambda}=2\tilde{b}$, the lowest curve of
Fig.~\ref{fig4}(a) and the left most curve of Fig.~\ref{fig4}(b)
correspond to the case of a single vortex. For the latter curve
the height of the barrier vanishes at
$\tilde{\Omega}_c\approx0.34$. (Notice that this corresponds to
the number obtained above using the Thomas-Fermi approximation,
namely $\tilde{\Omega}_c\approx0.5$). Thus, according to our
calculation, the critical frequency for vortex nucleation is
$\Omega_c/\omega_0=0.88$. This is to be compared with the
experimental number $\Omega_c/\omega_{\perp}=147/219=0.68$. Notice
that we consider $\omega_0$ to correspond to $\omega_{\perp}$
since in the experiment the axis of rotation is the $z$-axis.
Taking into account the approximations that we have used, we
consider this agreement to be reasonable. One possible reason for
the disagreement is that we have taken the core of the vortex to
be small compared to the distance over which the density of the
condensate changes significantly. This approximation is accurate
only in the deep region of the condensate and not close to the
Thomas-Fermi surface or away from it. Furthermore, our criterion
for vortex nucleation may be too restrictive since, even with a
nonzero barrier height, as shown in Fig.~\ref{fig4}(b), the vortex
can tunnel through the barrier and thus nucleate at a frequency
which is less than that obtained above.

\section{conclusion}
\label{conclusion} We have written down the energy functional of a
rotating Bose-Einstein condensate in terms of variables local to
its surface. Using a variational ansatz, we have calculated the
energy of a single vortex moving at the surface of the condensate.
This energy was used to investigate the energetic stability of the
vortex in terms of the rotation frequency of the confining
potential. Upon increasing the rotation frequency, the vortex
approaches the surface of the condensate. At a critical rotation
frequency, the vortex crosses the surface of the condensate and
enters it.

The local-density approximation was used to simplify the
calculations and lead to analytical results in the Thomas-Fermi
limit. This approximation is accurate only in the interior of the
condensate and not near the Thomas-Fermi surface or away from it.
The critical frequency obtained using this approximation is thus
not expected to be accurate. Therefore, we solve the problem
numerically including kinetic energy effects at the surface. This
leads to more accurate values for the critical frequency that
compare reasonably with experiment. It should be also noted that
our criterion for vortex nucleation, which is requiring that the
energy barrier vanish, may be too restrictive. When the height of
the energy barrier is nonzero but small enough, the vortex can
tunnel through it and thus nucleate at a rotation frequency that
is less than the one we calculated in the previous section. In
this sense, our calculated critical frequency may be regarded as
an upper bound to the vortex nucleation frequency.


\appendix

\section{Details of the energy Calculation}

In this Appendix we present details for the derivation of
Eq.~(\ref{energy2}) from the energy functional Eq.~(\ref{energy})
using the trial wave function Eq.~(\ref{psitrial}) for a moving
vortex.

Since the vortex axis is parallel to the $z$-axis and the vortex
is moving in the $y$-direction, the nontrivial contributions to
the energy arise only from integrating over $x$ and $y$.
Therefore, we calculate the energy of the vortex per unit length
of the $z$-direction, which means that we have to perform only
integrals in the $x$-$y$ plane. It is, thus convenient to shift
the origin of $\bf r^\prime$ from the surface of the condensate to
the core of the vortex and to express the integral of energy
functional in the polar coordinates $\rho$ and $\phi$. Explicitly,
we make the transformation ${\bf r^\prime}=x_0{\bf\hat
x}+{\bmath\rho}$.

The kinetic energy contribution reads
\begin{equation}
E_k=\int{ d{\bmath\rho}{\hbar^2\over2m} \left|\left(
{\bmath\nabla}-{im\over\hbar}{\bf\Omega}\times(x_0{\bf\hat x
}+{\bmath\rho})\right)\psi(\rho,\phi) \right|^2 }\;, \label{ke}
\end{equation}
where, here and through out this appendix, the gradient term is
with respect to $\bmath\rho$.
Inserting the trial function Eq.~(\ref{psitrial}) in this equation
we obtain
\begin{equation}
E_k\simeq {\hbar^2\over2m}n_0(x_0)\int{ d{\bmath\rho} \left|\left(
{\bmath\nabla}-{im\over\hbar}{\bf\Omega}\times(x_0{\bf\hat x
}+{\bmath\rho})\right)\chi(\rho)\exp{\left(i\Phi(\rho)\right)}
\right|^2 }\;. \label{ke2}
\end{equation}
Here, we have employed the approximation that the condensate
equilibrium density $n_0(x)$ is constant over distances of the
order of the coherence length. For this approximation to be valid,
the upper limit of the integration over $\rho$ must be much less
than the distance over which the condensate density changes
significantly. In this case we can approximate $n_0(x)$ by its
value at $x_0$ and take it out of the integration. Furthermore,
the last equation can be simplified to
\begin{eqnarray}
E_k&=& {\hbar^2\over2m}n_0(x_0)\int{ d{\bmath\rho} \left|
{\bmath\nabla}\chi+i\chi{\bmath\nabla}\Phi
-{im\over\hbar}{\bf\Omega}\times(x_0{\bf\hat x
}+{\bmath\rho})\chi\right|^2 }\nonumber\\
&=& {\hbar^2\over2m}n_0(x_0)\int{ d{\bmath\rho} \left[
({\bmath\nabla}\chi)^2 +(\chi{\bmath\nabla}\Phi)^2
+{m^2\over\hbar^2} \left({\bf\Omega}\times(x_0{\bf\hat x
}+{\bmath\rho})\right)^2\chi^2-2{m\over\hbar}
\left({\bf\Omega}\times(x_0{\bf\hat x
}+{\bmath\rho})\right)\cdot{\bmath\nabla}\Phi\right] }\;.
\label{ke3}
\end{eqnarray}
Knowing that ${\bmath\Omega}=\Omega{\bf\hat z}$ and using the
expression for $\Phi$ from Eq.~(\ref{phase}), it is
straightforward to show that the kinetic energy takes the form
\begin{equation}
E_k= {\hbar^2\over2m}n_0(x_0)\int{ d{\bmath\rho} \left[
({\bmath\nabla}\chi)^2+{\chi^2\over\rho^2}
+{m^2v_0^2\over\hbar^2}\chi^2
+{2m\over\hbar}\Omega(1-{m\over\hbar}x_0v_0)\chi^2
+{m^2\over\hbar^2}\Omega^2(x_0^2+\rho^2)\chi^2 \right] }\;.
\label{ke4}
\end{equation}

The contribution to the energy from the last term of
Eq.~(\ref{energy}) is
\begin{eqnarray}
E_\Omega&=& -{1\over2}m \,n_0(x_0)\int{ d{\bmath\rho} \left(
{\bf\Omega}\times(x_0{\bf\hat x}+{\bmath\rho})\right)^2\chi^2
}\nonumber\\
&=&
 -{1\over2}m\,\Omega^2\, n_0(x_0)\int{ d{\bmath\rho}
(x_0^2+\rho^2)\chi^2 } \;. \label{eo}
\end{eqnarray}
When added to the kinetic energy, this term cancels with the last
term of Eq.~(\ref{ke4}).

The contribution of the external trapping potential is given by
\begin{eqnarray}
E_{\rm ext}&=&F\,n_0(x_0)\int{d{\bmath\rho}(x_0+\rho\sin\phi)\chi^2}\nonumber\\
&=&F\,n_0(x_0)x_0\int{d{\bmath\rho}\,\chi^2}\;. \label{ev}
\end{eqnarray}

The contribution of the interatomic interaction reads simply
\begin{equation}
E_{\rm int}={1\over2}gn_0^2(x_0)\int{d{\bf\rho}\,\chi^4}\;.
\label{ie}
\end{equation}

Finally, we add all these contributions to obtain
Eq.~(\ref{energy2}).

\section*{Acknowledgments}
I would like to thank R.A. Duine, H.T.C. Stoof, H. Smith, C.J.
Pethick, and J.O. Andersen for fruitful discussions and helpful
suggestions.


\end{document}